\documentclass[10pt]{article}
\usepackage{latexsym,graphicx}
\newcommand{\be}{\begin{equation}}
\newcommand{\ee}{\end{equation}}
\def\n{\noindent}
\catcode `\@=11
\catcode `\@=12
\begin{document}
\begin{center}
\large{\bf {STRING COSMILOGICAL MODEL IN CYLINDRICALLY SYMMETRIC 
INHOMOGENEOUS UNIVERSE WITH ELECTROMAGNETIC FIELD II}} \\
\vspace{10mm}
\normalsize{ANIRUDH PRADHAN \footnote{Corresponding author}}\\
\normalsize{{\it Department of Mathematics, Hindu Post-graduate College, 
Zamania-232 331, Ghazipur, India \\
E-mail : pradhan@iucaa.ernet.in}} \\
\vspace{5mm}
\normalsize{SUNIL KUMAR SINGH, LAL JI SINGH YADAV}\\
\normalsize{{\it Department of Physics, S. D. J. Post-graduate College, 
Chandeswar-276 128, Azamgarh, India}} \\ 
\vspace{5mm}
\end{center}
\vspace{10mm}
\begin{abstract} 
Cylindrically symmetric inhomogeneous string cosmological model of the universe 
in presence of electromagnetic field is investigated. We have assumed that $F_{12}$ 
is the only non-vanishing component of electromagnetic field tensor $F_{ij}$. The 
Maxwell's equations show that $F_{12}$ is the function of $x$ alone whereas the magnetic 
permeability is the function of $x$ and $t$ both. To get the deterministic solution, 
it has been assumed that the expansion ($\theta$) in the model is proportional to 
the eigen value $\sigma^{1}~~_{1}$ of the shear tensor $\sigma^{i}~~_{j}$. Some physical 
and geometric properties of the model are also discussed.   
\end{abstract}
\smallskip
\n Keywords: cosmic string, electromagnetic field, inhomogeneous universe\\
\n PACS number: 98.80.Cq, 04.20.-q 
\section{INTRODUCTION}
Cosmic strings play an important role in the study of the early universe. These strings 
arise during the phase transition after the big bang explosion as the temperature goes 
down below some critical temperature as predicted by grand unified theories [1]-[5]. It 
is believed that cosmic strings give rise to density perturbations which lead to 
formation of galaxies [6]. These cosmic strings have stress energy and couple 
to the gravitational field. Therefore, it is interesting to study the gravitational effect 
which arises from strings. The general treatment of strings was initiated by Letelier 
[7, 8] and Stachel [9]. Magnetic fields are known to have a widespread presence in our 
Universe, being a common property of the intergalactic medium in galaxy clusters [10], 
while, reports on Faraday 
rotation imply significant magnetic fields in condensations at high red-shifts [11]. 
Studies of large-scale magnetic fields and their potential implications for the formation 
and the evolution of the observed structures, have been the subject of continuous 
investigation (see e.g. [12]-[19] for a representative though incomplete list). The origin 
of these fields, whether of astrophysical or cosmological origin, remains an unresolved 
issue. Also Harrison [20] has suggested that magnetic field could have a cosmological 
origin. If magnetism has a cosmological origin, as observation of $\mu G$ fields in galaxy 
clusters and high red-shift protogalaxies seem to suggest, it could have affected the 
evolution of the Universe [21]. As a natural consequences, we should include magnetic 
fields in the energy-momentum tensor of the early universe. The choice of anisotropic 
cosmological models in Einstein system of field equations leads to the cosmological models 
more general than Robertson-Walker model [22]. The presence of primordial magnetic fields 
in the early stages of the evolution of the universe has been discussed by several 
authors [23]-[31]. Melvin [32], in his cosmological solution for dust and electromagnetic 
field suggested that during the evolution of the universe, the matter was in a highly 
ionized state and was smoothly coupled with the field, subsequently forming neutral matter 
as a result of universe expansion. Hence the presence of magnetic field in string dust 
universe is not unrealistic. \\

Benerjee et al. [33] have investigated an axially symmetric Bianchi type I string dust 
cosmological model in presence and absence of magnetic field. The string cosmological 
models with a magnetic field are also discussed by Chakraborty [34], Tikekar and Patel 
[35, 36]. Patel and Maharaj [37] investigated stationary rotating world model with magnetic 
field. Ram and Singh [38] obtained some new exact solution of string cosmology with and 
without a source free magnetic field for Bianchi type I space-time in the different basic 
form considered by Carminati and McIntosh [39]. Singh and Singh [40] investigated string 
cosmological models with magnetic field in the context of space-time with $G_{3}$ 
symmetry. Singh [41, 42] has studied string cosmology with electromagnetic fields in 
Bianchi type-II, -VIII and -IX space-times. Lidsey, Wands and Copeland [43] have 
reviewed aspects of super string cosmology with the emphasis on the cosmological 
implications of duality symmetries in the theory. Bali et al. [44, 45] have 
investigated Bianchi type I magnetized string cosmological models.\\

Cylindrically symmetric space-time play an important role in the study of the universe 
on a scale in which anisotropy and inhomogeneity are not ignored. Inhomogeneous 
cylindrically symmetric cosmological models have significant contribution in 
understanding some essential features of the universe such as the formation of 
galaxies during the early stages of their evolution. Bali and Tyagi [17] and 
Pradhan et al. [18, 19] have investigated cylindrically symmetric inhomogeneous 
cosmological models in presence of electromagnetic field. Barrow and Kunze [46, 47] 
found a wide class of exact cylindrically symmetric flat and open inhomogeneous string 
universes. In their solutions all physical quantities depend on at most one space 
coordinate and the time. Recently Baysal et al. [48], Kilinc and Yavuz [49] have 
investigated some string cosmological models in cylindrically symmetric inhomogeneous 
universe. Kilinc and Yavuz [50] have also obtained string cosmology with magnetic field 
in cylindrically symmetric space-time. The case of cylindrical symmetry is natural 
because of the mathematical simplicity of the field equations whenever there exists 
a direction in which the pressure equal to energy density. \\

Motivated by the situation discussed above, in this paper, we have obtained 
a new solution of Einstein's field equations in cylindrically symmetric space-time in 
presence of electromagnetic field and string as a source. We have taken string and 
electromagnetic field together as the source gravitational field as magnetic field 
are anisotropic stress source and low strings are one of anisotropic stress source as 
well. This paper is organized as follows: The metric and field equations are presented 
in Section $2$. In Section $3$, we deal with the solution of the field equations in 
presence of electromagnetic field with perfect fluid distribution. In Section $4$, we deal 
with some physical and geometric properties of model. Concluding remarks are given in 
Section $5$. \\

\section{THE METRIC AND FIELD EQUATIONS}
We consider the metric in the form 
\begin{equation}
\label{eq1}
ds^{2} = A^{2}(dx^{2} - dt^{2}) + B^{2} dy^{2} + C^{2} dz^{2},
\end{equation}
where $A$, $B$ and $C$ are functions of $x$ and $t$.
The energy momentum tensor for the string with electromagnetic field 
has the form 
\begin{equation}
\label{eq2}
T^{j}_{i} = \rho u_{i}u^{j} - \lambda x_{i}x^{j} +  E^{j}_{i},
\end{equation}
where $u_{i}$ and $x_{i}$ satisfy conditions
\begin{equation}
\label{eq3}
u^{i} u_{i} = - x^{i} x_{i} = -1,
\end{equation}
and
\begin{equation}
\label{eq4}
u^{i} x_{i} = 0.
\end{equation}
Here $\rho$ being the rest energy density of the system of strings, $\lambda$ the 
tension density of the strings, $x^{i}$ is a unit space-like vector representing 
the direction of strings so that $x^{1} = 0 = x^{2} = x^{4}$ and $x^{3} \ne 0$, 
and $u^{i}$ is the four velocity vector satisfying the 
following conditions
\begin{equation}
\label{eq5}
g_{ij} u^{i} u^{j} = -1.
\end{equation}
In Eq. (\ref{eq2}), $E^{j}_{i}$ is the electromagnetic field given by Lichnerowicz \cite{ref51} 
\begin{equation}
\label{eq6}
E^{j}_{i} = \bar{\mu}\left[h_{l}h^{l}\left(u_{i}u^{j} + \frac{1}{2}g^{j}_{i}\right) 
- h_{i}h^{j}\right],
\end{equation}
where $\bar{\mu}$ is the magnetic permeability and $h_{i}$ the magnetic flux vector 
defined by
\begin{equation}
\label{eq7}
h_{i} = \frac{1}{\bar{\mu}} \, {^*}F_{ji} u^{j},
\end{equation}
where the dual electromagnetic field tensor $^{*}F_{ij}$ is defined by Synge \cite{ref52} 
\begin{equation}
\label{eq8}
^{*}F_{ij} = \frac{\sqrt{-g}}{2} \epsilon_{ijkl} F^{kl}.
\end{equation}
Here $F_{ij}$ is the electromagnetic field tensor and $\epsilon_{ijkl}$ is the Levi-Civita 
tensor density.

In the present scenario, the comoving coordinates are taken as 
\begin{equation}
\label{eq9}
u^{i} = \left(0, 0, 0, \frac{1}{A}\right). 
\end{equation}
We choose the direction of string parallel to z-axis so that
\begin{equation}
\label{eq10}
x^{i} = \left(0, 0, \frac{1}{A}, 0 \right). 
\end{equation}
We consider the current as flowing along the $z$-axis so that $F_{12}$ is the only 
non-vanishing component of $F_{ij}$. Maxwell's equations  
\begin{equation}
\label{eq11}
F_[ij;k] = 0,
\end{equation}
\begin{equation}
\label{eq12}
\left[\frac{1}{\bar{\mu}}F^{ij}\right]_{;j} = J^{i},
\end{equation}
require that $F_{12}$ is the function of x-alone and the magnetic permeability is the 
functions of $x$ and $t$ both. The semicolon represents a covariant differentiation. 

The Einstein's field equations (with $\frac{8\pi G}{c^{4}} = 1$) 
\begin{equation}
\label{eq13}
R^{j}_{i} - \frac{1}{2} R g^{j}_{i}  = - T^{j}_{i},
\end{equation}
for the line-element (\ref{eq1}) lead to the following system of equations:  
\[
\frac{B_{44}}{B} + \frac{C_{44}}{C} - \frac{A_{4}}{A}\left(\frac{B_{4}}{B} + 
\frac{C_{4}}{C}\right) - \frac{A_{1}}{A}\left(\frac{B_{1}}{B} + \frac{C_{1}}{C}\right) 
-\frac{B_{1}C_{1}}{BC}  + \frac{B_{4} C_{4}}{B C} 
\]
\begin{equation}
\label{eq14}
= \left[\lambda + \frac{F^{2}_{12}}{2\bar{\mu} A^{2} B^{2}} \right] A^{2},
\end{equation}
\begin{equation}
\label{eq15}
\left(\frac{A_{4}}{A}\right)_{4} - \left(\frac{A_{1}}{A}\right)_{1} + \frac{C_{44}}{C} - 
\frac{C_{11}}{ C} =  - \left[\frac{F^{2}_{12}}{2\bar{\mu} A^{2} B^{2}}\right] A^{2},
\end{equation}
\begin{equation}
\label{eq16}
\left(\frac{A_{4}}{A}\right)_{4} - \left(\frac{A_{1}}{A}\right)_{1} + \frac{B_{44}}{B} - 
\frac{B_{11}}{B} =   \left[\frac{F^{2}_{12}}{2\bar{\mu} A^{2} B^{2}}\right] A^{2},
\end{equation}
\[
- \frac{B_{11}}{B} - \frac{C_{11}}{C} + \frac{A_{1}}{A}\left(\frac{B_{1}}{B} + \frac{C_{1}}
{C}\right) + \frac{A_{4}}{A}\left(\frac{B_{4}}{B} + \frac{C_{4}}{C}\right) -
\frac{B_{1}C_{1}}{BC}  + \frac{B_{4} C_{4}}{B C} 
\]
\begin{equation}
\label{eq17}
= \left[\rho + \frac{F^{2}_{12}}{2\bar{\mu} A^{2} B^{2}}\right] A^{2},
\end{equation}
\begin{equation}
\label{eq18}
\frac{B_{14}}{B} + \frac{C_{14}}{C} - \frac{A_{4}}{A}\left(\frac{B_{1}}{B} + \frac{C_{1}}{C}
\right) - \frac{A_{1}}{A}\left(\frac{B_{4}}{B} + \frac{C_{4}}{C}\right) = 0,
\end{equation}
where the sub indices $1$ and $4$ in A, B, C and elsewhere denote ordinary differentiation
with respect to $x$ and $t$ respectively.

The velocity field $u^{i}$ is irrotational. The scalar expansion $\theta$, shear scalar 
$\sigma^{2}$, acceleration vector $\dot{u}_{i}$ and proper volume $V^{3}$ are respectively 
found to have the following expressions:
\begin{equation}
\label{eq19}
\theta = u^{i}_{;i} = \frac{1}{A}\left(\frac{A_{4}}{A} + \frac{B_{4}}{B} + \frac{C_{4}}{C}
\right),
\end{equation}
\begin{equation}
\label{eq20}
\sigma^{2} = \frac{1}{2} \sigma_{ij} \sigma^{ij} = \frac{1}{3}\theta^{2} - \frac{1}{A^{2}}
\left(\frac{A_{4}B_{4}}{AB} + \frac{B_{4}C_{4}}{BC} + \frac{C_{4}A_{4}}{CA}\right),
\end{equation}
\begin{equation}
\label{eq21}
\dot{u}_{i} = u_{i;j}u^{j} = \left(\frac{A_{1}}{A}, 0, 0, 0\right) 
\end{equation}
\begin{equation}
\label{eq22}
V^{3} = \sqrt{-g} = A^{2} B C,
\end{equation}
where $g$ is the determinant of the metric (\ref{eq1}). Using the field equations and 
the relations (\ref{eq19}) and (\ref{eq20}) one obtains the Raychaudhuri's equation as
\begin{equation}
\label{eq23}
\dot{\theta} = \dot{u}^{i}_{;i} - \frac{1}{3}\theta^{2} - 2 \sigma^{2} - \frac{1}{2} 
\rho_{p},
\end{equation}
where dot denotes differentiation with respect to $t$ and
\begin{equation}
\label{eq24}
R_{ij}u^{i}u^{j} = \frac{1}{2}\rho_{p}.
\end{equation}
 With the help of Eqs. (\ref{eq1}) - (\ref{eq4}), (\ref{eq9}) and (\ref{eq10}), the 
Bianchi identity $\left(T^{ij}_{;j}\right)$ reduced to two equations:
\begin{equation}
\label{eq25}
\rho_{4} - \frac{A_{4}}{A}\lambda + \left(\frac{A_{4}}{A} + \frac{B_{4}}{B} + 
\frac{C_{4}}{C}\right)\rho = 0
\end{equation}
and
\begin{equation}
\label{eq26}
\lambda_{1} - \frac{A_{1}}{A}\rho + \left(\frac{A_{1}}{A} + \frac{B_{1}}{B} + 
\frac{C_{1}}{C}\right)\lambda = 0.
\end{equation}
Thus, due to all the three (strong, weak and dominant) energy conditions, one finds 
$\rho \geq 0$ and $\rho_{p} \geq 0$, together with the fact that the sign of $\lambda$ 
is unrestricted, it may take values positive, negative or zero as well.  
\section{SOLUTION OF THE FIELD EQUATIONS}
As in the case of general-relativistic cosmologies, the introduction of inhomogeneities 
into the string cosmological equations produces a considerable increase in mathematical 
difficulty: non-linear partial differential equations must now be solved. In practice, 
this means that we must proceed either by means of approximations which render the non-
linearities tractable, or we must introduce particular symmetries into the metric of the 
space-time in order to reduce the number of degrees of freedom which the inhomogeneities 
can exploit. \\

To get a determinate solution, let us assume that expansion ($\theta$) in the model 
is proportional to the eigen value $\sigma^{1}~~_{1}$ of the shear tensor 
$\sigma^{i}~~_{j}$. This condition leads to
\begin{equation}
\label{eq27}
A = (BC)^{n},
\end{equation}
where $n$ is a constant. Equations (\ref{eq15}) and (\ref{eq16}) lead to
\begin{equation}
\label{eq28}
\frac{F^{2}_{12}}{\bar{\mu} B^{2}} = \frac{B_{44}}{B} - \frac{B_{11}}{B} - \frac{C_{44}}{C} 
+ \frac{C_{11}}{C}.
\end{equation}
and
\begin{equation}
\label{eq29}
2\left(\frac{A_{4}}{A}\right)_{4} - 2\left(\frac{A_{1}}{A}\right)_{1} + \frac{B_{44}}{B} - 
\frac{B_{11}}{B} + \frac{C_{44}}{C} - \frac{C_{11}}{ C} = 0.
\end{equation}
Using (\ref{eq27}) in (\ref{eq18}) reduces to
\begin{equation}
\label{eq30}
\frac{B_{41}}{B} + \frac{C_{41}}{C} - 2n \left(\frac{B_{4}}{B} + \frac{C_{4}}{C}\right)
\left(\frac{B_{1}}{B} + \frac{C_{1}}{C}\right) = 0.
\end{equation}
To get the deterministic solution, we assume 
\begin{equation}
\label{eq31}
B = f(x)g(t) ~ ~ \mbox{and} ~ ~ C = h(x) k(t)
\end{equation}
and discuss its consequences below in this paper.

In this case Eq. (\ref{eq30}) reduces to 
\begin{equation}
\label{eq32}
\frac{f_{1}/f}{h_{1}/h} = - \frac{(2n - 1)(k_{4}/k) + 2n(g_{4}/g)}{(2n - 1)(g_{4}/g) + 
2n(k_{4}/k)} = K \mbox{(constant)}.
\end{equation}
which leads to
\begin{equation}
\label{eq33}
\frac{f_{1}}{f} = K\frac{h_{1}}{h},
\end{equation}
and
\begin{equation}
\label{eq34}
\frac{k_{4}/k}{g_{4}/g} = \frac{K - 2nK - 2n}{2nK + 2n - 1} = a \mbox{(constant)}.
\end{equation}
From Eqs. (\ref{eq33}) and (\ref{eq34}), we obtain
\begin{equation}
\label{eq35}
f = \alpha h^{K},
\end{equation}
and
\begin{equation}
\label{eq36}
k = \delta g^{a},
\end{equation}
where $\alpha$ and $\delta$ are integrating constants.

From Eqs. (\ref{eq29}) and (\ref{eq27}), we obtain
\[
(2n + 1)\frac{B_{44}}{B} - 2n \frac{B^{2}_{4}}{B^{2}} + (2n + 1)\frac{C_{44}}{C} - 
2n\frac{C^{2}_{4}}{C^{2}} =
\]
\begin{equation}
\label{eq37}
(2n + 1)\frac{B_{11}}{B} + (2n + 1)\frac{C_{11}}{C} - 2n \frac{B^{2}_{1}}{B^{2}} - 2n \frac{C^{2}_{1}}
{C^{2}} = \mbox{N (constant)}.
\end{equation}
Eqs. (\ref{eq31}) and (\ref{eq37}) lead to
\begin{equation}
\label{eq38}
gg_{44} + r g^{2}_{4} = s g^{2},
\end{equation}
where 
$$
r = \frac{a(a - 1) - 2n(a + 1)}{(2n + 1)(a + 1)}, ~ ~ ~ s = \frac{N}{(2n + 1)(a + 1)}.
$$
Integrating Eq. (\ref{eq38}), we obtain
\begin{equation}
\label{eq39}
g = \left(c_{2} e^{bt} + c_{3} e^{-bt}\right)^{\frac{1}{(r + 1)}},
\end{equation}
where $b = \sqrt{s(1 + r)}$ and $c_{2}, c_{3}$ are constants of integration. Thus from (\ref{eq36}), 
we obtain 
\begin{equation}
\label{eq40}
k = \delta \left(c_{2} e^{bt} + c_{3} e^{-bt}\right)^{\frac{a}{(r + 1)}}.
\end{equation}
From Eqs. (\ref{eq33}) and (\ref{eq37}), we have
\begin{equation}
\label{eq41}
h h_{11} + \ell h_{1}^{2} = m h^{2},
\end{equation}
where
$$ \ell = \frac{K(K - 1) - 2n(K + 1)}{(2n + 1)(K + 1)},$$
$$ m = \frac{N}{(2n + 1)(K + 1)}.$$
Integrating (\ref{eq41}), we get
\begin{equation}
\label{eq42}
h = \left(r_{2} e^{\beta x} + r_{3} e^{-\beta x}\right)^{\frac{1}{(\ell + 1)}},
\end{equation}
where $\beta = \sqrt{m(\ell + 1)}$ and $r_{2}$, $r_{3}$ are constants of integration.   
Eqs. (\ref{eq35}) and (\ref{eq42}) lead to
\begin{equation}
\label{eq43}
f = \alpha \left(r_{2} e^{\beta x} + r_{3} e^{-\beta x}\right)^{\frac{K}{(\ell + 1)}}.
\end{equation}
Hence we obtain
\begin{equation}
\label{eq44}
B = f g = \alpha \left(r_{2} e^{\beta x} + r_{3} e^{-\beta x}\right)^{\frac{K}{(\ell + 1)}}
\left(c_{2} e^{bt} + c_{3} e^{-bt}\right)^{\frac{1}{(r + 1)}},
\end{equation}
\begin{equation}
\label{eq45}
C = \delta \left(r_{2} e^{\beta x} + r_{3} e^{-\beta x}\right)^{\frac{1}{(\ell + 1)}}
\left(c_{2} e^{bt} + c_{3} e^{-bt}\right)^{\frac{a}{(r + 1)}},
\end{equation}
\begin{equation}
\label{eq46}
A = (B C)^{n} = (\alpha \delta)^{n} \left(r_{2} e^{\beta x} + r_{3} e^{-\beta x}\right)^{\frac{n(K + 1)}
{(\ell + 1)}}\left(c_{2} e^{bt} + c_{3} e^{-bt}\right)^{\frac{n(a + 1)}{(r + 1)}}.
\end{equation}
Thus the geometry of the space-time (\ref{eq1}) reduces to the form
\[
ds^{2} =  (\alpha \delta)^{2n} \left(r_{2} e^{\beta x} + r_{3} e^{-\beta x}\right)^{\frac{2n(K + 1)}
{(\ell + 1)}}\left(c_{2} e^{bt} + c_{3} e^{-bt}\right)^{\frac{2n(a + 1)}{(r + 1)}} (dx^{2} - dt^{2}) \, +
\]
\[
\alpha^{2} \left(r_{2} e^{\beta x} + r_{3} e^{-\beta x}\right)^{\frac{2K}{(\ell + 1)}}
\left(c_{2} e^{bt} + c_{3} e^{-bt}\right)^{\frac{2}{(r + 1)}} dy^{2} \, +
\]
\begin{equation}
\label{eq47}
\delta^{2} \left(r_{2} e^{\beta x} + r_{3} e^{-\beta x}\right)^{\frac{2}{(\ell + 1)}}
\left(c_{2} e^{bt} + c_{3} e^{-bt}\right)^{\frac{2a}{(r + 1)}} dz^{2}.
\end{equation}
\section{SOME PHYSICAl AND GEOMETRIC PROPERTIES}
The energy density $(\rho)$, the string tension density $(\lambda)$, the particle 
density $(\rho_{p})$ for the model (\ref{eq47}) are given by 
\[
\rho = \frac{1}{(\alpha \delta)^{2n} \left(r_{2} e^{\beta x} + r_{3} e^{-\beta x}\right)^{\frac{2n(K + 1)}
{(\ell + 1)}}\left(c_{2} e^{bt} + c_{3} e^{-bt}\right)^{\frac{2n(a + 1)}{(r + 1)}}}\times
\]
\[
\Biggl[\frac{\beta^{2}\{n(K + 1)^{2} + K(\ell - K) + \ell\}}{(\ell + 1)^{2}}\frac{\left(r_{2} e^{\beta x} 
- r_{3} e^{-\beta x}\right)^{2}}{\left(r_{2} e^{\beta x} + r_{3} e^{-\beta x}\right)^{2}} \, +
\]
\[
\frac{b^{2}\{n(a + 1)^{2} + a\}}{(r + 1)}\frac{\left(c_{2} e^{bt} - c_{3} e^{-bt}\right)^{2}}
{\left(c_{2} e^{bt} + c_{3} e^{-bt}\right)^{2}} - \frac{\beta^{2}(K + 1)}{(\ell + 1)} \, - 
\]
\begin{equation}
\label{eq48}
\frac{F_{12}^{2}}{2\bar{\mu} \alpha^{2}\left(r_{2} e^{\beta x} + r_{3} e^{-\beta x}\right)^{\frac{2K}
{(\ell + 1)}} \left(c_{2} e^{bt} + c_{3} e^{-bt}\right)^{\frac{2}{(r + 1)}}}\Biggr],  
\end{equation}
\[
\lambda = \frac{1}{(\alpha \delta)^{2n} \left(r_{2} e^{\beta x} + r_{3} e^{-\beta x}\right)^{\frac{2n(K + 1)}
{(\ell + 1)}}\left(c_{2} e^{bt} + c_{3} e^{-bt}\right)^{\frac{2n(a + 1)}{(r + 1)}}}\times
\]
\[
\Biggl[\frac{b^{2}\{a(a - r) + r - n(a + 1)^{2}\}}{(r + 1)^{2}}\frac{\left(c_{2} e^{b t} 
- c_{3} e^{-b t}\right)^{2}}{\left(c_{2} e^{b t} + c_{3} e^{-b t}\right)^{2}} \, +
\]
\[
\frac{b^{2}(a + 1)}{(r + 1)} - \frac{\beta^{2}\{K(n + 1) + n\}}{(\ell + 1)}^{2}\frac{\left(r_{2} e^{\beta x} 
- r_{3} e^{- \beta x}\right)^{2}}{\left(r_{2} e^{\beta x} + r_{3} e^{-\beta x}\right)^{2}} \, -
\]
\begin{equation}
\label{eq49}
\frac{F_{12}^{2}}{2\bar{\mu} \alpha^{2}\left(r_{2} e^{\beta x} + r_{3} e^{-\beta x}\right)^{\frac{2K}
{(\ell + 1)}} \left(c_{2} e^{bt} + c_{3} e^{-bt}\right)^{\frac{2}{(r + 1)}}}\Biggr],  
\end{equation}
where
\[
F_{12}^{2} = \bar{\mu} \alpha^{2} \left(r_{2} e^{\beta x} + r_{3} e^{-\beta x}\right)^{\frac{2K}
{(\ell + 1)}}\left(c_{2} e^{bt} + c_{3} e^{-bt}\right)^{\frac{2}{(r + 1)}}\times
\]
\[
\Biggl[\frac{b^{2}(1 - a)}{(r + 1)}\left\{1 - \frac{(r - a)}{(r + 1)}\frac{\left(c_{2} e^{b t} 
- c_{3} e^{- b t}\right)^{2}}{\left(c_{2} e^{b t} + c_{3} e^{- b t}\right)^{2}}\right\} \, +
\]
\begin{equation}
\label{eq50}  
\frac{\beta^{2}(1 - K)}{(\ell + 1)}\left\{1 - \frac{(\ell - K)}{(\ell + 1)}\frac{\left(r_{2} e^{\beta x} 
- r_{3} e^{- \beta x}\right)^{2}}{\left(r_{2} e^{\beta x} + r_{3} e^{-\beta x}\right)^{2}}\right\}\Biggr].
\end{equation}
\[
\rho_{p} = \frac{1}{(\alpha \delta)^{2n} \left(r_{2} e^{\beta x} + r_{3} e^{-\beta x}\right)^{\frac{2n(K + 1)}
{(\ell + 1)}}\left(c_{2} e^{bt} + c_{3} e^{-bt}\right)^{\frac{2n(a + 1)}{(r + 1)}}}\times
\]
\[
\Biggl[\frac{\beta^{2}\{n(K + 1)(K + 2) + K(\ell - K) + (\ell + n)\}}{(\ell + 1)^{2}}\frac{\left(r_{2} 
e^{\beta x} - r_{3} e^{-\beta x}\right)^{2}}{\left(r_{2} e^{\beta x} + r_{3} e^{-\beta x}\right)^{2}} 
\]
\begin{equation}
\label{eq51}
+ \, \frac{b^{2}(a - r)(1 - a)}{(r + 1)^{2}}\frac{\left(c_{2} e^{bt} - c_{3} e^{-bt}\right)^{2}}
{\left(c_{2} e^{bt} + c_{3} e^{-bt}\right)^{2}} - \frac{\beta^{2}(K + 1)}{(\ell + 1)} + \frac{b^{2}(a + 1)}
{(r + 1)}\Biggr].  
\end{equation}
The scalar of expansion $(\theta)$, shear tensor $(\sigma)$, acceleration vector $\dot{u}_{i}$ and the 
proper volume for $(V^{3})$ for the model (\ref{eq47}) are given by 
\begin{equation}
\label{eq52}
\theta = \frac{(n + 1)(a + 1)b\left(c_{2} e^{bt} - c_{3} e^{-bt}\right)}{(r + 1)(\alpha \delta)^{n} 
\left(r_{2} e^{\beta x} + r_{3} e^{-\beta x}\right)^{\frac{n(K + 1)}{(\ell + 1)}}\left(c_{2} e^{bt} 
+ c_{3} e^{-bt}\right)^{\frac{n(a + 1)}{(r + 1)} + 1}},
\end{equation}
\begin{equation}
\label{eq53}
\sigma^{2} = \frac{b^{2}\{(a + 1)^{2}(n^{2} - n + 1) - 3a\}\left(c_{2} e^{bt} - c_{3} e^{-bt}\right)^{2}} 
{3(r + 1)^{2}(\alpha \delta)^{2n} \left(r_{2} e^{\beta x} + r_{3} e^{-\beta x}\right)^{\frac{2n(K + 1)}
{(\ell + 1)}}\left(c_{2} e^{bt} + c_{3} e^{-bt}\right)^{\frac{2n(a + 1)}{(r + 1)} + 2}}
\end{equation}
\begin{equation}
\label{eq54}
\dot{u}_{i} = \left(\frac{n(K + 1)\beta}{(\ell + 1)}\frac{\left(r_{2} e^{\beta x} - r_{3} e^{- \beta x}\right)}
{\left(r_{2} e^{\beta x} + r_{3} e^{-\beta x}\right)}, 0, 0, 0 \right),
\end{equation}
\begin{equation}
\label{eq55}
V^{3} = (\alpha \delta)^{2n + 1} \left(r_{2} e^{\beta x} + r_{3} e^{-\beta x}\right)^{\frac{(2n + 1)(K + 1)}
{(\ell + 1)}}\left(c_{2} e^{bt} + c_{3} e^{-bt}\right)^{\frac{(2n + 1)(a + 1)}{(r + 1)}}.
\end{equation}
From Eqs. (\ref{eq52}) and (\ref{eq53}), we obtain 
\begin{equation}
\label{eq56}
\frac{\sigma^{2}}{\theta^{2}} = \frac{(a + 1)^{2}(n^{2} - n + 1) - 3a}{3(n + 1)^{2}(a + 1)^{2}} = 
\mbox {constant}.
\end{equation}
The model (\ref{eq47}) represents an expanding, shearing and non-rotating 
universe. The expansion in the model increases as time increases when $n < 0$ 
but the expansion in the model decreases as time increases when $n > 0$. The 
spatial volume $V$ increases as time increases. If we set the suitable values of 
constants, we find that energy conditions $\rho \geq 0$, $\rho_{p} \geq 0$ are 
satisfied. We observe that $\frac{\sigma}{\theta}$ is constant throughout. 
Hence the model does not approach isotropy. The energy density $(\rho)$ and string tension density 
$(\lambda)$ decrease as electromagnetic field component $(F_{12})$ increases. The 
electromagnetic field tensor $(F_{12})$ becomes in uniform state when $x = 0$, $t = 0$ and 
it increases when $x$ and $t$ increase. \\
\section{CONCLUDING REMARKS}
In this paper, we have investigated the behaviour of a string in the cylindrically 
symmetric inhomogeneous universe in presence of electromagnetic field 
with perfect fluid distribution. Generally the model is expanding, shearing, non-rotating 
and accelerating. The solution obtained in this paper is new and different from the other 
author's solutions. In this solution all physical and kinematical quantities depend on at 
most one space coordinate and the time.\\

\section*{ACKNOWLEDGEMENTS} 
One of the authors (A. P.) would like to thank Professor G. Date, IMSc., Chennai, India 
for providing facility where the part of this work was carried out.
  

\end{document}